\documentclass[epj]{svjour}

\usepackage{amsmath}	
\usepackage{color}
\usepackage{mathtools, cuted}
\usepackage{hyperref}
\pdfoutput=1
\newcommand{\braket}[1]{\langle {#1} \rangle }
\newcommand{\ket}[1]{|{#1} \rangle }

\title{Radioactive beams and inverse kinematics: probing the quantal
	texture of the nuclear vacuum }
\author{F. Barranco \inst{1} \and G. Potel \inst{2} \and E. Vigezzi \inst{3} \and R. A. Broglia \inst{4,}\inst{5}}
\institute{Departamento de F\`isica Aplicada III,
	Escuela Superior de Ingenieros, Universidad de Sevilla, Camino de los Descubrimientos, 	 41092 Sevilla, Spain \and National Superconducting Cyclotron Laboratory, Michigan State University, East Lansing, MI 48824, USA \and INFN Sezione di Milano, via Celoria 16, I-20133 Milano, Italy \and The Niels Bohr Institute, University of Copenhagen, 
	DK-2100 Copenhagen, Denmark \and Dipartimento di Fisica, Universit\`a degli Studi di Milano,
	Via Celoria 16, 
	I-20133 Milano, Italy}
\begin{abstract}
	{The properties of the quantum electrodynamic (QED) vacuum in general, and of the nuclear vacuum (ground) state 
	in particular are determined by virtual processes implying the excitation of a photon and of an electron--positron pair in the first case
	and of, for example,    the excitation of a collective quadrupole surface  vibration and a  particle--hole pair
	in the nuclear case.  
	Signals of these processes can be detected in the laboratory in terms of what can be considered 
	a nuclear analogue of Hawking radiation. An analogy which extends to other physical processes involving QED vacuum fluctuations 
	like the Lamb shift, pair creation by $\gamma-$rays, van der Waals forces and the Casimir effect,
	to the extent that one concentrates on the eventual outcome 
	resulting  by forcing  a virtual process  to become real,  and not on  the role of 
	the black hole role in defining the event horizon.  In the nuclear 
	case, the role of this event is taken over at a microscopic, fully quantum mechanical level, by   
	nuclear probes (reactions) acting on a virtual particle of the zero point fluctuation (ZPF) of the nuclear vacuum in a similar irreversible, no--return,  fashion as the event horizon does, letting the other particle, entangled with the first one, escape to infinity, and eventually be detected. With this proviso in mind one can posit that the reactions 
	$^1$H($^{11}$Be,$^{10}$Be$(2^+$;3.37 ${\rm MeV}$))$^2$H and
	$^{1}$H($^{11}$Li,$^9$Li($1/2^-$; 2.69 ${\rm MeV}$))$^3$H together with the associated $\gamma-$decay processes
	indicate a possible nuclear analogy 	of Hawking radiation.} 	
\end{abstract}

\begin{document}

 \maketitle
\section{ Introduction}

At the basis of quantum mechanics one finds Heisenberg's indeterminacy relations, Born--Jordan
commutation rules, Pauli principle, Born probability interpretation of Schr\"o\-dinger
wave function and Dirac transformation theory. 
All these  elements find natural imagery in Feynman diagrams, and call for the existence 
of a vacuum state, whose structure is determined by virtual processes. 
These processes, which do not  conserve energy, modulate the quantum  vacuum through the 
transient presence of fermionic particles  and antiparticles and of  bosonic quanta. 
In the case of the electromagnetic vacuum 
permeating space, these are virtual off--shell 
electron--positron pairs and photons 
(Fig. \ref{fig1}(I)(a)).
If some of these elements are  modified through the action of an external field 
which provides in the process energy, angular and linear momentum, etc., the remaining particles can become
on-shell and thus the vacuum radiates. Reaching the detector, this radiation provides  information on the virtual states,
and thus on the texture of the vacuum, let alone on the event triggering the radiation. 

Let us build  the case  one step at a time,  and start considering pair creation 
in the laboratory  by a photon (left wavy line and vertex, Fig. \ref{fig2}(a)). Because the created pair 
has invariant finite mass, while the photon has zero mass, a second interaction is necessary. In Fig. \ref{fig2}(a),
it is provided by a second photon (lower right wavy line) and associated vertex, reflecting the action of a massive charge $Z$ (cross 
labeled Z) needed for momentum conservation. It is of notice that in all vertices one finds three particles. 
This is in keeping with the fact that in QED the interaction acting at each vertex is bilinear in fermions
(electrons, positrons) and linear in bosons (photons) (see App. A). Within this context, the process associated 
with the vertex to the left in Fig. \ref{fig2}(a) plays the same role  as the lower vertex of Fig. \ref{fig1}(a)(I). 
Returning to Fig. \ref{fig2} (a), if one  allows 
the electron to annihilate with the positron (closing the loop) and absorb the photon (left wavy line), one 
obtains the Feynman diagram of Fig. \ref{fig1}(a)(I), as the presence 
of the massive charge $Z$ and associated photon is not needed. 

By collapsing the two vertices and the massive charge of the QED Feynman diagram of Fig. \ref{fig2} (a)  into
a  single vertex assumed to result
from the action of the curved gravitational space associated with a black hole (cylinder), one can 
adapt \ref{fig2} (a) to Hawking's heuristic diagram  (Fig. \ref{fig2} (b); see \cite{Hawking:75,Hawking:77}). We return to this point below, but before let us consider
the possibility to subject  the QED vacuum to a supercritical atomic nucleus  of effective charge $Z > Z{crit} \approx 180$,
resulting from a  quasimolecular state  transiently formed in a heavy ion collision. Under such conditions 
the QED vacuum state is expected to become charged, positrons being emitted at the same time. 
The vacuum rearranges in such a way so a to minimise  the effect of the applied "external" field. That is, 
the vacuum acts as a screening medium. A schematic representation of such a process is given in Fig. 2(c). The heavy grey
lines provide a schematic representation of the two ions  at the distance of closest approach, of the order of 16 fm. 
The transient, quasimolecular state of charge $Z=Z_1+Z_2 > Z_{crit}$ leads to a multi photon process of pair creation \cite{Muller:72,Rafelski:74,Soff:77}.

As in the black hole case, the $Z> Z_{crit}$ situation shows  a preferred distance (radius) for the occurrence 
of the phenomenon leading to particle emission. In the simplest black hole description it is the Schwarzchild radius,
while in the heavy ion collision leading to $Z_{crit}$ it is that of the radius of the 
1s orbital with $\epsilon_{s_{1/2}} < -2 m_e c^2$ of the quasimolecular system.

In a similar way in which Hawking's approach  \cite{Hawking:75} is based on a classical (relativistic) gravitational picture, the calculation
of the heavy ion collision eventually leading to the $Z_1+ Z_2 = Z_{cr}$ is described 
in a time-dependent semiclassical approximation up to the distance of closest approach.
But from   that point on,  the pair production  and associated radiation  process is carried out quantum mechanically, as testified
by the QED Feynman diagram (c) of Fig. \ref{fig2} . Within this context it is in principle thinkable to follow a similar 
approach in dealing with Hawking radiation and instead of (b) Fig. \ref{fig2}, use a diagram similar to (c). It is of notice that while the $Z_{crit}$  heavy ion reaction positron production phenomenon did not lead, in spite of much experimental effort,  to a conclusive answer \cite{Rafelski:16}, a similar "shake off" phenomenon of the QED
	vacuum  was observed  in the process $\omega + n \omega_0 \to e^+ e^-$ for $n \geq 4$ laser photons of wavelength 527 nm colliding with
	a photon energy of 29 GeV \cite{Burke:97}. Within this context see also \cite{Dumlu:10,Yu:19} and references therein.

In the above scenario nothing precludes the possibility of the presence of both photons and gravitons in the corresponding Feynman diagrams. 
In fact this seems to be the most likely situation, as shown in \cite{Parentani:99} in connection 
with photon emission from a charged particle falling into a black hole, described  within the framework of QED and found 
to be of Hawking type, although some nontrivial differences with 
the ``classical'' result were found. 

QED vacuum fluctuations play also a central role in the Lamb shift 
\cite{Lamb:47,Bethe:47,Welton:48,Kroll:49,French:49}, as seen from 
Figs. \ref{fig1} (II)(b)-(d) and Fig. \ref{fig3} (C) (see also \cite{Weinberg:96}). The corresponding self-energy processes 
depend on the atomic orbital occupied by the electron. In the case of hydrogen it leads to a splitting between the $^2P_{1/2}$ and 
$^2S_{1/2}$ orbital of 1058 MHz. Within this context, we mention that Hawking refers to the Lamb shift as a phenomenon which provides
confirmation of virtual fluctuations of QED \cite{Hawking:77}. As mentioned  above,
this phenomenon  implies   processes involving photons,
aside from electrons and positrons  (Figs.  \ref{fig1} (II) (b)-(d)) and Fig. \ref{fig3}). The fact that Hawking states that quantum mechanics  implies that 
the {\it whole} of space (and not only that close to the event  horizon of black holes) is filled with pairs of "virtual" particles and antiparticles 
that are constantly  materialising in pairs must imply that he views the process displayed in Fig. \ref{fig1} (a) 
(right, i.e. with a photon) equivalent to that  shown on the left, where the photon is represented  by the associated electromagnetic field. 

Let us now briefly mention the Casimir effect \cite{Casimir:48,Casimir:48b,Casimir:48c,Jaffe:05}, 
originally intended to provide a quantum mechanical description of the van der Waals 
force \cite{London:30,London:37,Lifschitz:55,Bjorken:98,Israelachvili:85,Pauling:63} acting between  two non-polar molecules, taking into account retardation effects. 
The results  obtained correspond to the
long--wavelength limit of the Feynman diagram shown in Fig. \ref{fig4} (d), and  connected with vacuum ZPF. In other words, 
while it is true that  the Casimir energy  can be expressed in terms of Feynman diagrams with external legs \cite{Jaffe:05}, this does not mean that 
they are  not a direct consequence of QED vacuum zero point fluctuations (within this context see Figs. \ref{fig1} (II)(b) and (d) and \ref{fig4} (a)-\ref{fig4} (d)).
The Casimir effect is referred to, if not by name, quite explicitly on p.202 of \cite{Hawking:75}, in connection with the statement that 
the black hole being an excited state of the gravitational field can decay quantum mechanically and that because of quantal fluctuations, energy 
should be able to tunnel out of the corresponding potential well, a particle creation analogous  to that caused by a deep potential 
well in flat space  (confinement of two infinite walls) \cite{Bjorken:98}. 
It is of notice that  a detailed determination of the Casimir effect requires surface plasmons  to be  considered \cite{Intravaia:07}. Within 
the nuclear connection it is closely connected with induced nuclear interaction, in particular induced pairing interaction 
\cite{Barranco:99,Terasaki:02a,Brink:05}.

\section{ Hawking radiation}

After the above has been  stated, we can use Fig. \ref{fig1} I (b) in what follows. We start with the left hand side representation of electromagnetic quantum fluctuations (Fig. \ref{fig1} I (a)), 
for then  use the QED description (right hand side Feynman diagram).

At the zero point fluctuation (ZPF) domain centered around $r_0$, all particles can, in principle,  become real  
(on shell). 
Adopting a simple Newtonian description, this can happen in the case of the electron-positron pair, provided 
\begin{align}
\nonumber &(m_e c^2 + T_{e^-} + (U_G)_{e^- -bh}) \\
&+(m_e c^2 + T_{e^+}  + (U_G)_{e^+-bh}) = 0,
\end{align}
where $(U_G)_{i-bh} $
%= -G (m_{bh} \times m_i/r)_{r_0}$ 
is the 
gravitational interaction energy between the black hole ($bh$, of mass $m_{bh}$) and the 
 electron and  positron ($e^-,e^+,m_e)$. 
%The gravitational constant is denoted $G$, while 
The quantities $T_{e^-}$
and $T_{e^+}$ are the kinetic energies associated with these  fermions.
%$h \nu_0$ being the photon energy. 
The functions $U_G(r)$  vanish at $r=\infty$, the remaining quantities 
being all positive. Thus,  both particles cannot be emitted together as $HR$. But if one of them 
falls behind the event horizon characterised by the Schwarzchild radius 
($r_s = 2 G m_{bh} /c^2)$, 
%in which case $h \nu_S + (U_G(r_S)_{ph-nh} =0$) 
and the associated gravitational energy  $(U_G)_{i-bh}$ is sufficiently negative, the on-shell 
condition can eventually be fulfilled and HR emitted.

Assuming the electron escapes to infinity ($(U_G)_{e^--bh}$=0) with kinetic energy $T_{e^-}$, the trapped
(infalling) positron-black hole gravitational interaction provides the negative energy necessary to fulfill global
energy conservation. Because the subsystem ($bh+e^+$)  has less energy than the original $bh$, one can posit
that the $bh$ has lost mass which has been emitted as an electron (HR).

Let us now consider the case in which the pair materialises through an elementary QED vacuum fluctuation 
(Fig. \ref{fig1} (I)(b) right hand side diagram).
The above equation should then  include the photon energy and associated gravitational interaction with the $bh$ field,
\begin{align}
\nonumber &\left(h \nu_0 + (U_G)_{ph-bh}\right) + (m_e c^2 +T_{e^-} + (U_G)_{e^--bh})\\& + (m_e c^2 + T_{e^+}+(U_G)_{e^+-bh} )= 0, 
\end{align}
where $h \nu_0$ is the photon energy while $m_{ph} = h \nu/c^2$ is the photon mass entering $(U_G)_{ph-bh}$.
Being three the particles present in the vacuum ZPF (Fig. \ref{fig1} I (a) right), a variety of escape combinations are possible. It is sensible to think that a $bh$ 
radiates as a black body. Thus, all possible particles combinations as well as final-state interactions  are expected to
be present, similarly to what happens in QED pair creation  by a supercritical Coulomb field (Fig. \ref{fig2} (c)).  Because not only gravitons can induce pair production, but also
photons in presence of the $bh$, the situation resembles that of the "classical" result for the Casimir force per unit area 
between two parallel plates separated by a distance $d$ ($F_C = - \hbar c \pi^2 /(240 \times  d^4))$ \cite{Lifschitz:55}. In fact, this expression
is only valid in the limit $\alpha \to 0$ of the fine structure constant, and assuming perfect conductivity. As mentioned above, 
the QED expression of the Van der Waals force between two metallic plates depends on the corresponding surface plasma, let alone on $\alpha$
\cite{Jaffe:05,Intravaia:07}.

In the case in which the photon escapes as $HR$ (Fig. \ref{fig1} (I)(b)), it will eventually be observed that the original 
emission frequency undergoes a strong gravitational red shift.
Being emitted near the event horizon, the asymptotic ($r = \infty$) frequency 
is 
\begin{equation}
h\nu_{\infty} = h \nu_0 \left(1 - \frac{r_s}{r_0} \right)^{1/2} = h \nu_0 + (U_G(r_0))_{ph-bh} + ... 
\end{equation}

Regarding the process in which the photon associated with the vacuum ZPF falls behind the horizon,
 $E_{HR} = 2 m_e c^2 + T_{e^-} + T_{e^+}$ (Fig. \ref{fig1} (I)(c)).
 Assuming electron-positron pair annihilation takes place (in presence of the massive object 
 which ensures linear momentum conservation) , may lead to photon production of frequency
  $\nu' \geq 2m_e c^2 /h$.

Predicted more than forty years ago \cite{Hawking:75}, Hawking radiation through which 
black holes  lose energy and mass, eventually evaporating 
(primordial black holes), still awaits experimental confirmation. 
In fact, it is difficult if not impossible to observe Hawking radiation from a real black hole, (see however
\cite{An:18} )  and analogue black--hole 
experiments are being studied in search for alternative examples  of it  (cf. \cite{Steinhauer:16,Castelvecchi:16}
and refs. therein).

\section{ Nuclear  Field Theory: structure  and reactions of exotic nuclei}
 
Quantum electrodynamics (QED) in Feynman  formulation provides a detailed description 
of the electromagnetic vacuum, paradigm of the quantum  vacuum  \cite{Schweber:94}.
Nuclear field theory (NFT), tailored after Feynman's graphical version of QED, supplemented by
renormalization, allows for  a quantal description of nuclear structure in general and of the 
nuclear vacuum in particular  \cite{Bes:74,Bortignon:77,Bes:77c,Broglia:16}.

In this description particles ($p$) and holes ($h$), namely nucleons moving above and missing from the Fermi sea respectively, play the role of electrons and positrons. They are to be calculated as solutions of the Hartree-Fock mean field. Collective vibrations, play the role of photons. The strength of the particle-vibration coupling vertices play the role of the fine structure constant. Such vertices are to be summed up to infinite order to calculate the vibrations which, at variance to the photon, are composite modes. A further difference is that the ``nuclear photons'' come in a number of species, namely of $ph$-type (e.g. surface vibrations) of, $pp$-type (pairing vibrations), as well as a variety of spin, isospin, etc. quantum numbers.

Worked out in the seventies in connection with nuclear structure NFT has been further developed to systematically deal with spontaneously broken symmetries and associated phase transitions and Goldstone modes \cite{Bes:90}, and generalized to deal, on equal footing, with structure and reactions \cite{Broglia:04a,Broglia:75}. Making use of renormalization techniques, convergence in non--perturbative situations can be ensured \cite{Broglia:16}. NFT has been applied to deal with a wide variety of phenomena throughout the mass table, providing an overall account of the experimental findings at the 10\% level \cite{Barranco:17,Barranco:19}, and predictions which tested, were found in accordance with observations at a similar level of accuracy \cite{Barranco:01,Tanihata:08,Potel:10}.

Because of its graphical rules, NFT allows to make parallels and find unexpected connections with many--body theories of condensed matter and cluster physics \cite{Mahaux:85,Broglia:04b}, let alone QED. In particular, in connection with analogues to the Lamb shift in the systematic probing of the nuclear quantum vacuum (ground state) (see Fig. \ref{fig3} and e.g. \cite{Barranco:17}). This is the reason why it appears natural to elaborate on a possible parallel between nuclear phenomena (structure and reactions), and Hawking radiation. Connection extended to the Casimir effect triggered by the remark found on p. 202 of \cite{Hawking:75}, namely: ``This particle creation is directly analogous to that caused by a deep potential well in flat space--time''. In other words, pair production of QED vacuum under stress (constrain). 

In this connection we also note that the realistic description of the Casimir effect involves the consideration of the fluctuations of the QED vacuum (exchange of virtual photons). Generalizing these phenomena to the dynamical Casimir effect (conducting plates in relative acceleration) the connection with HR through Einstein's equivalence principle emerges in a natural fashion \cite{Nation:12,Nugayev:87}. Given the parallel existing between NFT and QED, replacing the moving plates by the colliding nuclei in a nuclear reaction, the nuclear analogue of HR seems permissible.

Reactions using  exotic radioactive nuclei in inverse kinematics and active cell targets setups
have brought the study of  the nuclear structure and reactions to unexpected heights and technical 
refinements. This is mainly a consequence  of the efforts made to achieve a complete description of the nuclei
under study, reflected in the use of a wide variety of probes leading to Coulomb excitation and 
inelastic scattering and associated $\gamma-$decay, as well as inducing one- and two-nucleon transfer 
reactions. This is particularly so 
in the probing of nuclei lying at the edge of matter stability as is the case of neutron drip line 
systems.
Paradigmatic examples of such developments are studies carried out at TRIUMPH  \cite{Tanihata:13}, Saclay and GANIL
\cite{Keeley:04} and RIKEN  \cite{Motobayashi:12}, which have provided, among other things,
detailed information on the vacuum state of exotic nuclei. The reason for concentrating our attention on these nuclei is because, being weakly bound and close to the neutron drip line, they display very large fluctuations.

\begin{figure*}[h]
	\begin{center}
		%\fbox{\include graphics[width=0.7\textwidth]{fig10.pdf}}
		\includegraphics[width=13cm]{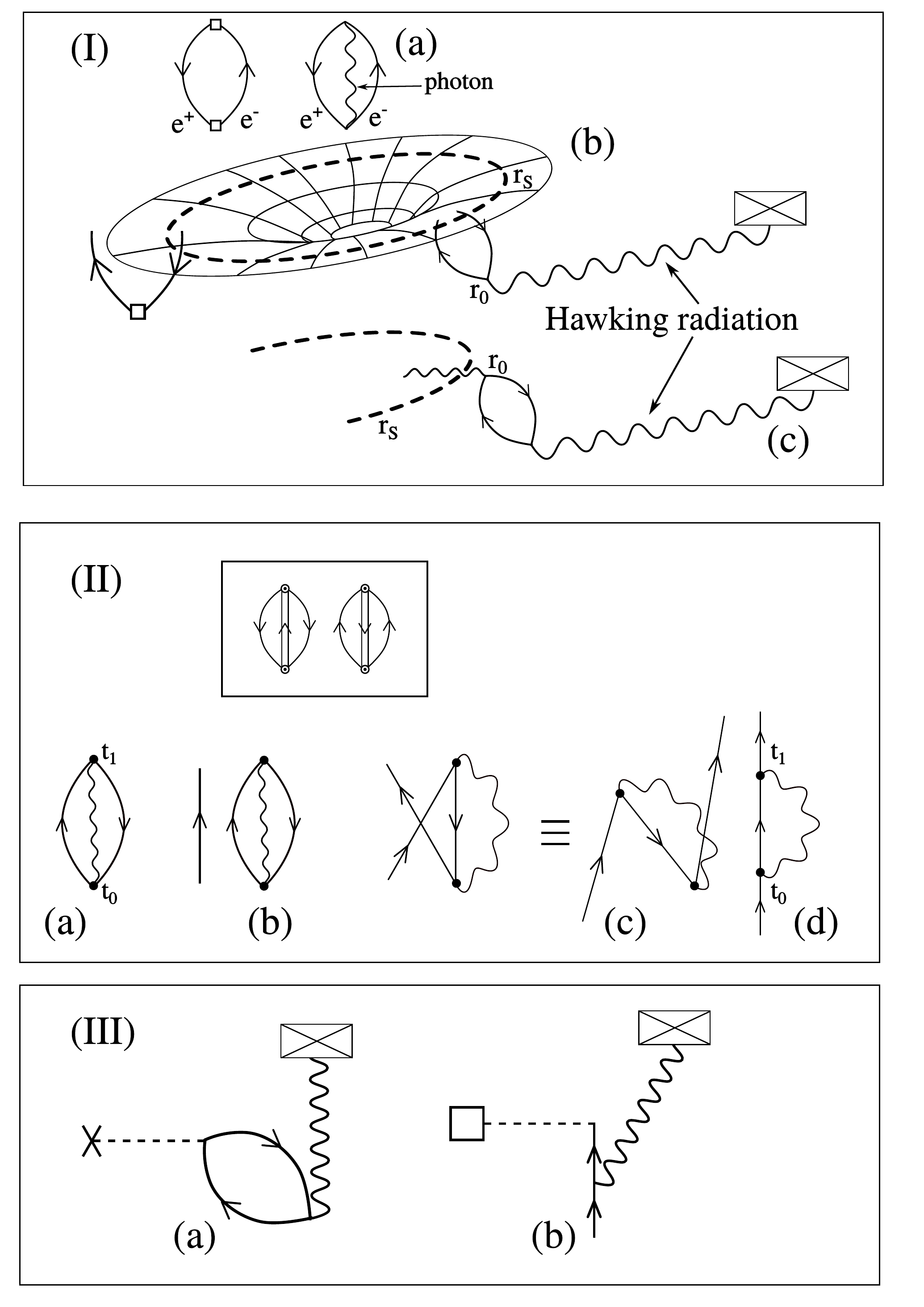}
	\end{center}
	\caption{
		{\bf  (I) (a)}  ZPF of the QED vacuum produce  virtual photons and electron-positron pairs.  The virtual photon can
		be represented by the associated electromagnetic potential (empty square, left diagram, see also (b) left) or by a wavy line (right diagram)
		(see e.g. Figs. 8.11.1(a)  and 8.11.2(b) of ref.  \cite{Schweber:94}).
		When
		{\bf (b)}   the pair (  {\bf (c)}  or the photon) gets trapped behind the event horizon of a black hole, the point beyond 
		which the gravitational pull is too strong even for light to escape, before the virtual process closes, the virtual photon
		(pair) become real. The photon (pair, eventually recombining in a photon) that escapes 
		is emitted as  Hawking radiation, which eventually can be recorded  as a click 
		in a detector (crossed rectangle).
		{\bf (II)(a)}  ZPF  of the vacuum (ground) state of an even nuclear system 
		(upward  (downward) arrowed line  representing a nucleon (nucleon hole); 
		wavy line, a  vibration);
		%associated with  closed shell nucleus, e.g. the $N=6$ isotones $^{10}$Be and $^9$Li. particle (hole) 
		% are represented as arrowed lines pointing upwards (downwards), while collective 
		%  $(p,h)$ vibrations are described as wavy lines. 
		{\bf (b)} odd system;  {\bf (c)} Pauli principle correction
		between the particle considered explicitly and those involved in the vibration (Lamb shift like diagram);
		{\bf (d)}    time ordering of the previous process. In the inset we show ZPF associated  with addition and subtraction pairing vibrational
		modes (double arrowed lines).
		{\bf (III)(a) }  Acting with an external field (cross + dashed line, inelastic scattering e.g. (p,p')) on process (a) of (II) before it closes,
		that is at a time $t_0 < t <t_1$, one can force the virtual fluctuations of the vacuum
		to become real, and eventually observe a click in the detector (crossed rectangle). {\bf (b) }  Similar 
		information is obtained by intervening process (d) of (II) with the appropriate external field 
		(empty square + dashed line, e.g. (p,d) reaction ).    }\label{fig1} 
\end{figure*}

\begin{figure*} 
	\begin{center} 
		\includegraphics[width=12cm]{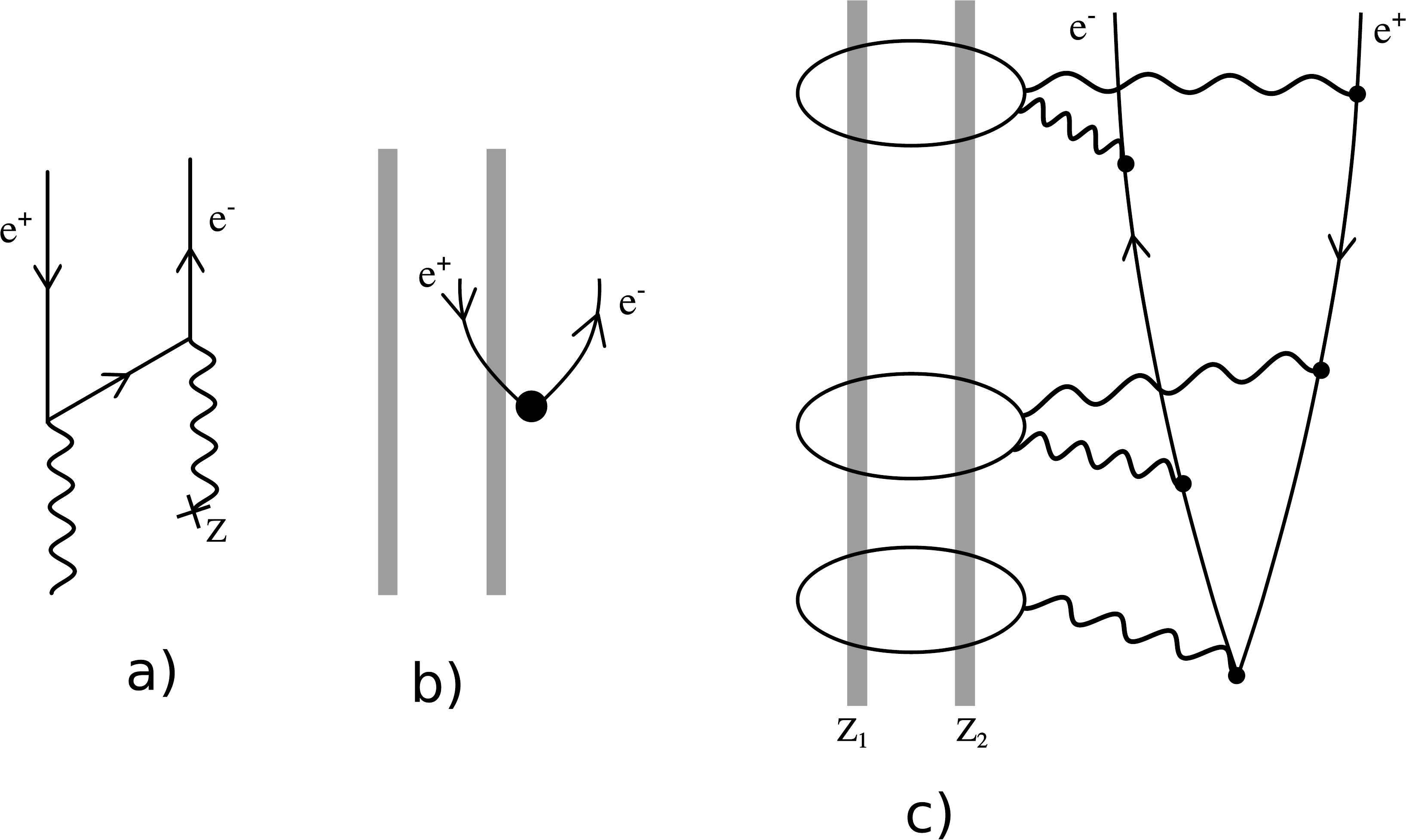}
	\end{center}
	\caption{Schematic representation of {\bf (a)} Pair creation (arrowed lines pointing upwards (downwards) represent 
		electrons (positrons), while wavy lines describe photons); (
		{\bf (b)} Hawking radiation, the hatched circle represents 
		the left photon in (a)  and the two vertices, the massive charge not being 
		needed in presence of a black hole (Schwarzchild
		radius, hatched cylinder); {\bf (c)} Transient quasimolecular state  of two heavy ions at the distance of closest approach 
		($Z= Z_1 + Z_2 > Z_{crit} \approx $ 180) making the QED vacuum unstable. 
	}\label{fig2}
\end{figure*}

\begin{figure*} 
 	\begin{center} 
 		\includegraphics[width=0.8\textwidth]{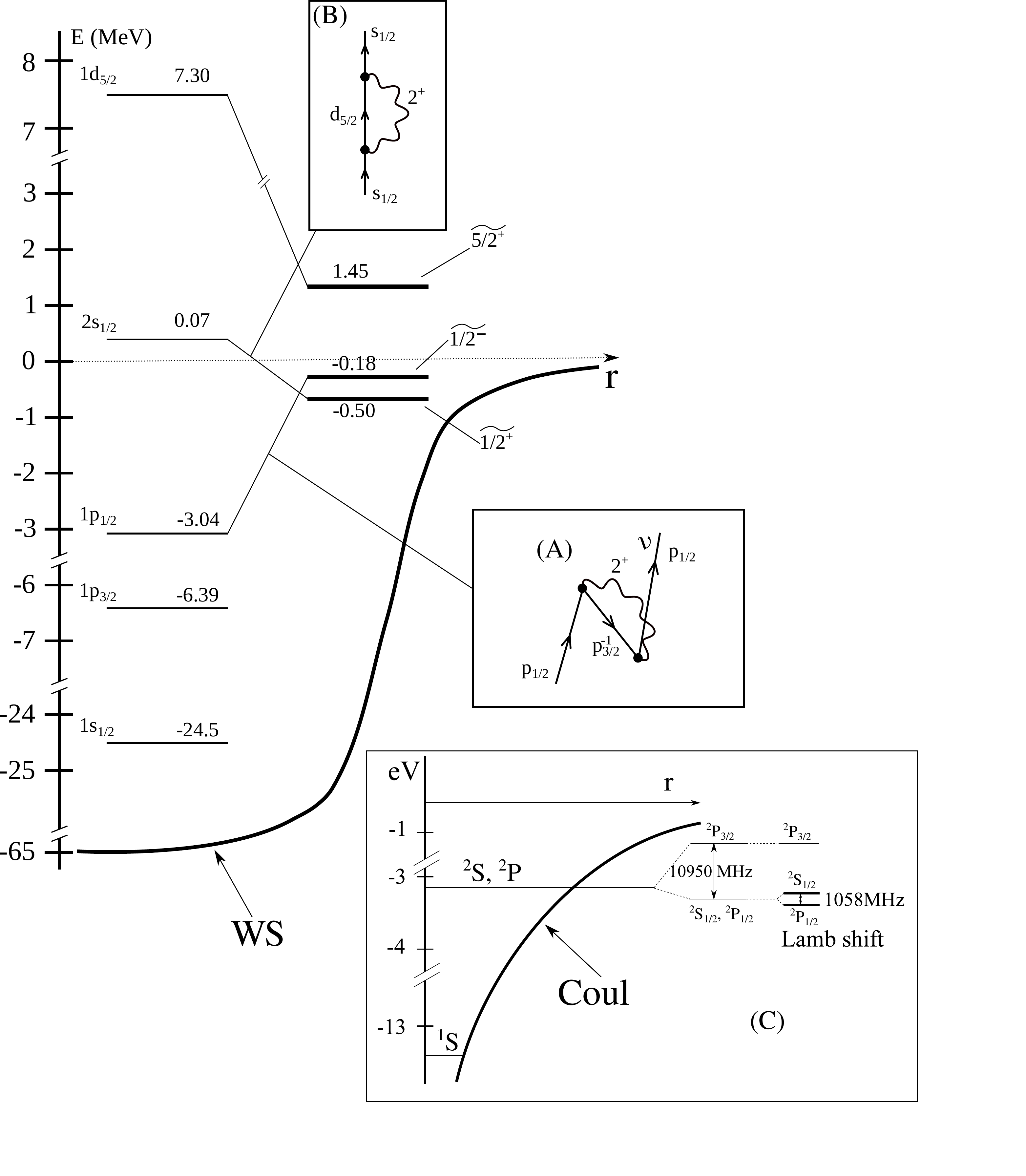}
 	\end{center}
 	\caption{Bare (thin) and dressed (bold face horizontal lines)
 		single-particle levels of $^{11}$Be calculated using a Woods-Saxon (WS) mean field. 
 		Due to the dressing of neutron motion with mainly quadrupole vibrations
 		of the core $^{10}$Be (insets (A) and (B)) inversion in sequence between the 
 		$ 2s_{1/2}$ and $ 1p_{1/2}$ levels  (parity inversion) is observed. The numbers 
 		are   energies in MeV. In inset (C), the lowest energy levels of hydrogen atom
 		%(Coulomb field (Coul) 
 		are indicated, the Coulomb potential (Coul) is also schematically shown.
 		The effects of fine structure according to Dirac theory ($l,s$ coupling  plus relativistic
 		mass increase)  and Lamb shift associated with the splitting of the 
 		$^2S_{1/2}$ and $^2P_{1/2}$ levels are displayed.}\label{fig3}
 \end{figure*}

\begin{figure*} 
 	\begin{center} 
 		\includegraphics[width=0.8\textwidth]{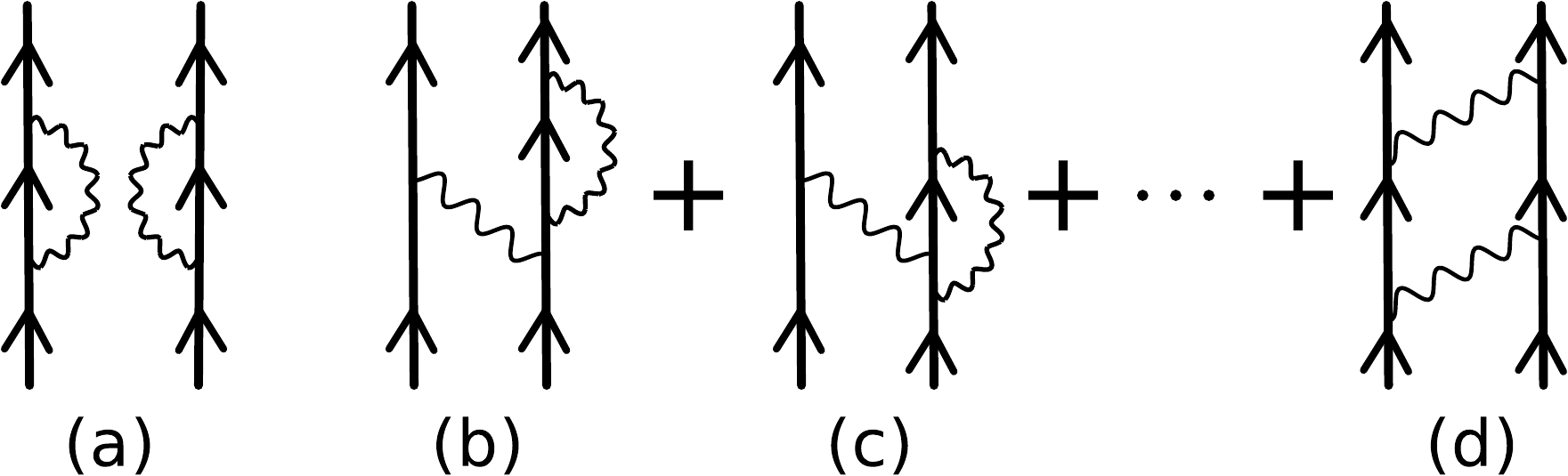}
 	\end{center}
 	\caption{{\bf (a)} Self energy diagrams of two electrons (nucleons) e.g. of two hydrogen atom
 		electrons (closed shell nuclei plus one nucleon), arrowed lines indicating the fermions, wavy-lines 
 		photons (nuclear particle-hole (ph) -like vibrations); {\bf(b)} Coulomb interaction (induced nuclear interaction) 
 		resulting from the exchange of a photon ((ph)-vibration); {\bf (c)} same as above but vertex corrected; 
 		{\bf (d)} Van der Waals interaction (higher order induced nuclear interaction). }\label{fig4}
 \end{figure*}

\begin{figure*}
 	\centering
 	%\fbox{\includegraphics[width=0.7\textwidth]{Fig10.pdf}}
 	%\begin{minipage}{0.4\textwidth}
 	\includegraphics[width=11cm]{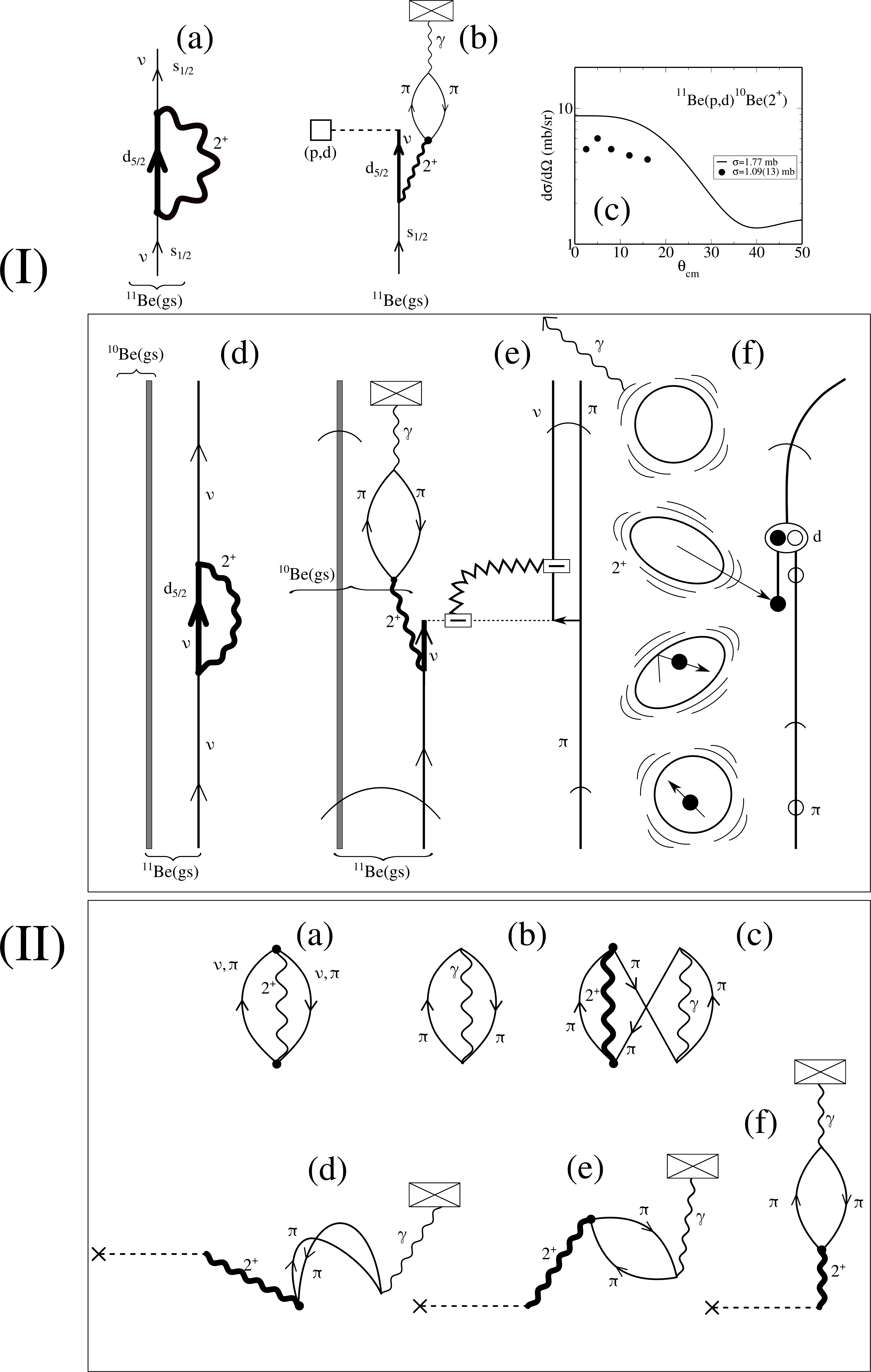}
 	\caption{ {\bf (I)} A virtual process in nuclear physics becomes real through the action of an external field. Arrowed lines pointing upwards (downwards) indicate a nucleon ($p$) (nucleon hole ($h$)) while wavy lines labeled $2^+$ denote a quadrupole collective ($ph$) vibration. Bold face line stand for fully dressed modes (renormalization procedure \cite{Broglia:16})
 		{\bf (a)}  Clothing process of the $1/2^+$ parity inverted ground state of $^{11}_4$Be$_{7}$
 		through the coupling to the low-lying quadrupole vibration of the core $^{10}_4$Be$_6$; 
 		%the  detailed structure of the NFT diagram is displayed in ${\bf (d)}$;  
 		{\bf (b)} schematic representation 
 		of the pickup of the neutron moving around a $N=6$ closed shell and populating the low-lying
 		quadrupole vibrational state of this core, in coincidence with the corresponding $\gamma-$decay 
 		(see also {\bf II (f)}); the structure and reaction NFT diagram describing the pickup
 		process in inverse kinematics, i.e. $^1$H($^{11}$Be,$^{10}$Be($2^+$,3.368 MeV))$^2$He is 
 		shown in {\bf (d)} and  {\bf (e)} together with a cartoon representation in {\bf (f)} (the jagged line represents 
 		a graphic mnemonic of the recoil effect, see \cite{Broglia:16} App. F as well as \cite{Potel:13} , App. A).
 		Proton 
 		and neutrons are labeled $\pi$ and $\nu$ respectively, while $d$ stands for deuteron.
 		Curved arrows indicate projectile motion (reaction). Normal arrowed 
 		lines, motion inside target or projectile (structure).
 		{\bf (c)} predicted (continuous curve) and experimental
 		(solid dots) absolute differential cross sections     associated with the indicated  pickup process. 
 		{\bf (II)} Interaction of protons in a nucleus with nuclear vibrations (solid dot, PVC vertex 
 		$\beta_L R_0  \partial U/\partial r Y_{LM}^*(\hat r)$ \cite{Bohr:75}, $\beta_L$: dynamical
 		distortion parameter, $U(r)$ central potential) and photons (normal vertex, see also Fig. \ref{fig1} (I)(a), 
 		electromagnetic 
 		interaction $e \int d^4 x J_{\mu}(x) A^{\mu} (x)$,$A^{\mu}$ being 
 		the vector potential, and $J_{\mu}$ the current density ($\mu=1,...,4)$ \cite{Holstein:89}).
 		While the variety of diagrams shown have general validity, we have assumed
 		we are dealing with the low-lying correlated particle-hole quadrupole vibration 
 		($L=2)$ of $^{10}_4$Be$_6$ lying at 3.368 MeV,  $B(E2; 0^+ \to 2^+$) = 0.0052 $e^2 b^2$ being
 		associated with $\beta_2 \approx 0.9$. An arrowed line pointing  upward (downward) 
 		describes a proton (proton hole) moving in the $p_{1/2}$ ($1p_{3/2}$) orbital. 
 		Zero point fluctuations of the nuclear ground state associated with : {\bf (a)} the nuclear
 		vibration, {\bf (b)} the electromagnetic field associated with the corresponding spontaneous
 		$\gamma-$decay.   {\bf (c)} Pauli principle correction to the simultaneous presence 
 		of the above two ZPF processes. {\bf (d)} Intervening the virtual excitation of the nuclear vibrations 
 		(graph (c)) with an external (inelastic) field  (cross followed by a dashed line), in coincidence with the $\gamma-$decay 
 		($\gamma-$detector, crossed box), the virtual process (c) becomes real. {\bf (e),(f)}  time ordering of the 
 		above process correspond to the  RPA contributions through backwardsgoing and forwardsgoing amplitudes
 		\cite{Bohr:75} and subsequent $\gamma-$decay.
 	} \label{fig5}
 	%\end{minipage}
 \end{figure*}

The zero point fluctuations associated with $^{11}$Be and $^{11}$Li  cores
(see Fig. \ref{fig1} (II)(a) as well as boxed inset), self energy contribution of the parity inverted (Fig. \ref{fig3} )
ground state $1/2^+$  of $^{11}$Be (Fig. \ref{fig1} (II) (d)), see also Fig. \ref{fig5} (I) (a)) and of the induced pairing 
correlation of the halo neutrons of $^{11}$Li  (Figs. \ref{fig6} (b) where also a 1$^-$ vibration is to be
considered in this last case), 
contribute   approximately 
6.3\% and 4.7\% of the corresponding binding energies, respectively  (Table 1). A major fraction of the associated ZPF  mass defect
in these nuclei is contributed by processes which involve the quadrupole modes:
 86\% in the case of $^{11}$Be and 74\% in that of $^{11}$Li 
(for details of the general framework see e.g. \cite{Baroni:04} and refs. therein).

Direct experimental insight into the mechanism at the basis of the above results in particular, 
and of the structure properties of the two halo nuclei in question  can be obtained through 
one- and two-nucleon processes, namely \cite{Winfield:01} 
$^{1}$H($^{11}$Be,$^{10}$Be(($2^+$;3.368 MeV))$^2$H and
% \footnote{It is of notice that the probabilities of populating the final excited states
%of $^9$Li through channels alternative to the direct ones are considerably smaller, leading to cross
%sections  which are three orders of magnitude smaller than experimentally observed, and can be neglected
%(see Table 1 of ref. \cite{Potel}). A similar situation is expected in the case of the
%population of the first $2^+$ excited state of $^{10}$Be \cite{Winfield}.} 
\cite{Tanihata:08} \\$^{1}$H($^{11}$Li,$^9$Li($1/2^-$;2.69 MeV))$^3$H.

  The mass
relations, which parallel (2)  are in these cases (see Fig. \ref{fig5} (I)(b),(d)--(f) in relation to the first 
reaction and Fig. \ref{fig6} (c) in connection with the second one),
\begin{equation}\label{eq4}
M(_ZX_f)c^2 = M(_ZX_i) c^2  - h \nu_{2^+} + ((\Delta m)c^2 + \Delta T),
\end{equation}
with

\begin{equation} 
\Delta m = 
\begin{cases} 
(m_p -m_d) \quad [ ((_ZX_i,\, _ZX_f) \equiv (^{11}_4\text{Be},\,_4^{10}\text{Be}))], \\
 (m_p- m_t) \quad [((_ZX_i,\, _ZX_f)  \equiv (^{11}_3\text{Li},\,^{9}_3\text{Li}))],
\end{cases} 
\end{equation} 

and
\begin{equation} 
\Delta T = 
\begin{cases} 
(T(^{11}\text{Be}) - T(^{10}\text{Be}) - T_d, \quad [ (Z=4)], \\
T(^{11}\text{Li}) - T(^9\text{Li}) - T_t, \quad [(Z=3)],
\end{cases}
\end{equation} 
where $p,d,$ and $t$ label proton, deuteron and triton respectively. The  term inside parentheses
in the left hand side of (3) takes into account the kinetic energy of the projectile inducing the 
nuclear reaction and of the resulting outgoing particles, the other term being associated 
with the reaction $Q-$value.
 Although the outcome of the $\gamma-$coincidence experiment (related to the $h\nu_{2^+}$ term in (4)) can be taken
 for granted,  its actual measurement in processes based on inverse kinematics like the 
 ones under consideration, is technically quite trying and has not yet been measured. Be as it may, the fact that the calculated absolute transfer differential
  cross sections provide an overall account of the experimental findings \cite{Barranco:17,Potel:10} gives direct 
  insight into the soundness, the (renormalised) NFT picture of the 
 nuclear vacuum state, has (see Fig. \ref{fig5} (I)(c) and lower inset of Fig. \ref{fig6}).
 
 Let us now return to Fig. \ref{fig1} (II). The bare properties of an odd nucleon moving around the core (Fig. \ref{fig1} (II)(b)) get modified though 
 Pauli principle corrections (Fig. \ref{fig1} (II)(c)) and through the associated dressing  process
 resulting from its time ordering (Fig. \ref{fig1} (II)(d)).  Within the scenario of quantum electrodynamics (QED) where Feynman diagrams 
 were developed, and in keeping with the symmetry existing between positron and electron phase spaces, {\bf N}-like and 
 self-energy-like  \cite{Schweber:94} processes
 (Figs. \ref{fig1} (II)(c) and (d)) are operative on equal footing. Observation of any of the associated virtual processes 
 dressing the electron by interrupting it through the action of an external field (e.g. Fig. \ref{fig1} (III)(b)), carries similar information concerning 
 both contributions II(c) and (d).  Because of spatial quantisation, finite nuclei display an asymmetry between occupied and 
 empty states (particles and holes). As a consequence process (c) of Fig. \ref{fig1} (II) may be allowed and  not  process (d),  or viceversa.
 This is particularly true for light nuclei, for example $^{11}$Be \cite{Barranco:17}.

 In the core of $^{11}$Be, namely $^{10}_{4}$Be$_{6}$, six neutrons occupy the $1s_{1/2}$ and 1$p_{3/2}$ levels (Fig. 3). The 
 dominant ZPF is of quadrupole type, the main neutron  component being associated with the 
 $((p_{1/2},p^{-1}_{3/2})\otimes 2^+)_{0^+}$ ZPF (Fig. 5(II)(a)). Because $\epsilon_{p1/2} -
 \epsilon_{p3/2}\approx 3.38 $ MeV and $\hbar \omega_{2^+} =$3.368 MeV, the largest 
 amplitude of the wavefunction of the quadrupole mode is associated with  the  neutron particle-hole excitation $(p_{1/2},p^{-1}_{3/2})_{2^+}$.
 The repulsion due to Pauli principle correction  (Fig. \ref{fig3}  inset (A))is $\approx 2.86$ MeV.  
 The clothing of the $2s_{1/2}$ bare level by the quadrupole mode
 (Fig. \ref{fig3} inset (B))
 %see also Fig. 3(I)(a)) 
 makes it  heavier, 
 lowering its energy by about 0.5 MeV (570 keV). The result of the two processes mentioned above 
 is parity inversion, and the  appearance of  the $N=6$ new magic number together with the melting 
 away of the $N=8$ standard one.
In a similar way  in which the Lamb shift  (Fig. \ref{fig3}, inset C)  provides a measure 
of  the fluctuations of the QED vacuum
(see \cite{Pais:86}, p. 451),  parity inversion 
measures ZPF of the nuclear vacuum (ground) state. In this last case further information 
can be obtained as compared with the atomic case, through particle transfer reactions. 

Let us elaborate on this point.
Interpreting  the arrowed lines of Fig. \ref{fig1} (III)(a) as an  electron and a positron, the wavy curve as 
a photon and the external field (cross + dashed line) as the event horizon 
of a black hole (see Fig. \ref{fig1} (I)(b)) one has a Feynman representation of Hawking radiation. A nuclear analogue 
of such radiation, to the extent that one considers only the wavy line and the detector click, 
is provided by graph (b) of Fig. \ref{fig1} (III), if one interprets the arrowed line as a nucleon,
the wavy line as a nuclear vibration and the external field (open square+  dashed line) as a irreversible and nucleon pickup reaction
intervening the self--energy process shown in Fig. \ref{fig1} (II)(d)  at a time $t$ fulfilling $t_0 < t < t_1$.
A concrete example of the above parlance  is provided by the no return event corresponding to the one neutron pickup reaction 
of the single-halo valence nucleon of $^{11}$Be, leading to the population of the low-lying 
quadrupole, first excited  (vibrational) state of the core $^{10}$Be, as  shown in Figs. \ref{fig5} (I)(b),(e) 
and (c) (see also
Fig. \ref{fig5} (II) in relation with the spontaneous 
\footnote{Spontaneous $\gamma$-decay is a direct consequence of the ZPF of  the nuclear 
vacuum (through its proton component) due to the presence of the ZPF of the electromagnetic field.
} 
$\gamma-$decay of the $2^+$ state, in
coincidence with the reaction process).

Light nuclei at the drip line provide another paradigmatic example of parity inversion 
and of a nuclear analogue, again in the sense of a virtual process becoming real 
through the irreversible action of an external field.  The nucleus is $^{11}$Li, the no return event  
in question the process    $^{1}$H($^{11}$Li,$^9$Li($1/2^-$;2.69 MeV))$^3$H  (Fig. \ref{fig6} (c)). 
The $\ket{^{11}\text{Li(gs)}}$  can be viewed as  a two-neutron halo pair 
addition mode (double arrowed line) and a proton moving in the $p_{3/2}$ orbital which acts  as a spectator.
In Figs. \ref{fig6} (a) and (b),  virtual processes associated with self-energy and induced pairing interaction (vertex
corrections)
are shown (for details see ref. \cite{Barranco:01}). Acting with an external two-nucleon pickup field at a time $t$
such that $t_0 < t < t_1$, leads to the population of the     $\ket{^{11}\text{Li}(1/2^-;2.69\text{ MeV})}$ first excited state
with an absolute differential cross section (Fig. \ref{fig6}, lower part) accounting for the experimental findings (see ref. \cite{Tanihata:08}).

  \begin{table*}[h!]
\begin{center}
%\begin{adjustbox}{max width=\textwidth}
\begin{tabular}{|c  | c |  c | c |  c |}
\hline 
& $BE/A$(keV) &  $BE_{ZPF}$ (keV) & $BE$(keV)  &  $BE_{ZPF}/BE$ ($\times 10^{-2}$)  \\ \hline
$^{11}$Be         & 5952.54$\pm 22$ & 4110&  65477.94  &  6.3\\ \hline
$^{11}$Li           & 4155.38$\pm 6$  & 2150 & 45709.18 & 4.7\\ \hline 
\end{tabular}
%\end{adjustbox}
\end{center}
\caption{ Binding energy $BE$ and binding energy per nucleon $BE/A$ (both in keV)
(see  \cite{Bachelet:08} and Suppl. material of this reference)
associated with
 the nuclei  $^{11}$Be  and $^{11}$Li. The ZPF contribution to the binding energy   
  of these two nuclei (see ref. \cite{Baroni:04} for details of the general framework) arises 
 from dipole, quadrupole and octupole $(p,h)$ 
 vibrations and from monopole pair vibrational modes (see also \cite{Potel:14}).}
\label{table3}
\end{table*}

\section{Entanglement and correlations}
The characteristic trait of quantum mechanics is the fact that when two systems, of which one knows the states through their respective wavefunctions enter into temporary physical interaction, and when after a time of mutual influence the systems separate again, then they can no longer be described in the same way as before, as they have become entangled. After restablishing one wavefunction by observation, the other one can be inferred simultaneously \cite{Schrodinger:35}. With the proviso that detector sensitivity is adequate to cope with background noise, let alone set up to pick up the specific signal of the phenomenon under study. A macroscopic manifestation of quantum entanglement is provided by superconductivity in bulk metals at low temperature, and by  Josephson current through an unbiased junction.
The Josephson effect provides a macroscopic manifestation of quantum entanglement. But to detect the supercurrent circulating through an unbiased junction between two weakly coupled superconductors it is necessary to go from standard 100$\Omega$ junctions easy to operate with, to 1$\Omega$ ones, let alone eliminate the earth magnetic field, as well as to carry out quantitative investigations to distinguish the effect from tiny superconducting shorts \cite{Anderson:64b,McMillan:69}. 

Within this context, arguably, is it possible to set in the proper perspective the failure to detect the QED vacuum instability through collisions between very heavy ions. This is in keeping with the complexity of calculating absolute cross sections in such cases \cite{Broglia:04a}, let alone analyze experiments associated with highly excited, massive nuclei which eventually can convert their many--body energy into pairs $(e^-,e^+)$ \cite{Rafelski:16}. At variance, in the case of direct reactions, in particular those under discussion $((p,d),\,(p,t))$, carried out at moderate bombarding energies (3 MeV/$A$), only few channels and elementary modes of excitation are open and active respectively. Furthermore one, in these cases, knows how to calculate absolute cross sections which reproduce the experimental findings within a 10\% error.     

\begin{figure*}
 	\begin{center}
 		%\fbox{\includegraphics[width=0.7\textwidth]{fig10.pdf}}
 		\includegraphics[width=18cm]{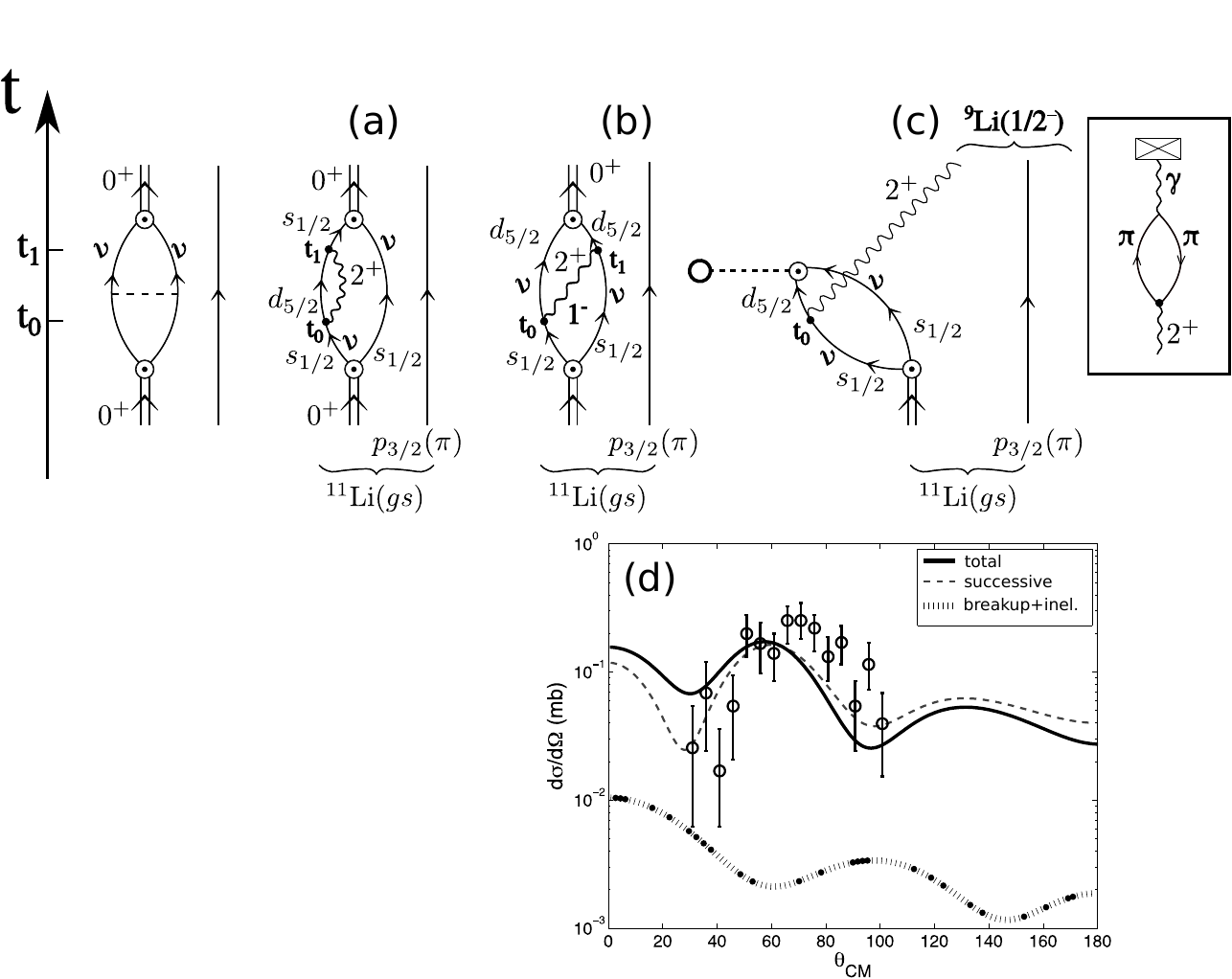}
 	\end{center}
 	\caption{{\bf (a),(b)} Virtual process associated with the correlation
 		of the two halo neutrons of  $_3^{11}$Li$_8$ moving around the closed shell $N=6$
 		core $_3^9$Li$_6$ due to the bare $NN$-$^1S_0$ interaction and to the exchange of collective vibrations, as well as the dressing of the halo neutrons. The odd $p_{3/2}$ proton ($\pi$) is assumed to act as a spectator;
 		{\bf (c)} an external two-neutron pickup field (open circle + dashed line) as provided by the 
 		inverse kinematic reaction $^1$H($^{11}$Li,$^9$Li($1/2^-$))$^3$H populates the lowest excited
 		state of $^9$Li, $1/2^-$ member of the multiplet
 		$(p_{3/2}(\pi) \otimes 2^+)$, forcing the $2^+$ vibration of the  core 
 		to become on shell and eventually by coupling to the electromagnetic field (see upper right boxed inset), $\gamma-$decay (HR) (further aspects of the physics associated with the reaction process are elaborated 
 		in connection with HR, see Fig. \ref{fig5} (II); \textbf{(d)}: the theoretical absolute differential cross section (continuous curve)
 		is compared with the data (open circles with error bars) in the lower box inset (see \cite{Tanihata:08} and \cite{Potel:10}). }\label{fig6}
 \end{figure*}

Within this context it is of notice that the probabilities of populating the final state $1/2^-$ in the reaction $^{11}$Li$(p,t)^9$Li($1/2^-$; 2.691 MeV) through channels alternative to the direct, one--step ones, are considerably less important. They lead to cross sections which are three orders of magnitude smaller than experimentally observed (see Fig. \ref{fig6} (d), and Table I of ref. \cite{Potel:10}). A similar situation is expected in the case of the population of the first $2^+$ excited state of $^{11}$Be in the reaction $^{11}$Be$(p,d)^{10}$Be($2^+$; 3.368 MeV) \cite{Winfield:01}. Concerning entanglement of the escaping Hawking particle (detected $\gamma$--ray) with its partner(s) swallowed in the black hole (picked up in the no--return reaction process), the nuclear examples under discussion are amenable to a technically trying, but straightforward control, known as coincidence experiments. Namely, to accept events in which the photon ($\hbar\nu_{2^+}$=3.368 MeV) and the deuteron (Fig \ref{fig5} (e)) or the photon  ($\hbar\nu_{2^+}$=2.691 MeV) and the triton (Fig. \ref{fig6} (c)) are recorded gating the corresponding detectors at the energy $\hbar\nu_{2^+}$ and at that resulting from (\ref{eq4}) respectively.  Entanglement which extends over the physical dimensions of modern RIB laboratory detector setups.

An alternative, simpler experiment, which carries equal \textit{bona fide} quantum mechanical entanglement information but is arguably less technically demanding is the following. Identify only the nature of the outgoing particle, or set up a $\gamma$--detector array to record a single line of frequency $\nu_{2^+_1}$ and intensity $I_\gamma\pm a$. The quantity $I_\gamma$ is related to the absolute transfer cross section and $a$ to the associated experimental error. In this way one eliminates any possible contributions from other channels but the direct one (see e.g. Fig. \ref{fig6} (d)).

From a quantum mechanical point of view, once the click in the $\gamma$--detector has disentangled the outgoing ($2^+_1$ vibration$\to$ ($\gamma$--decay), see Fig. \ref{fig6} (d)) particle wavefunction $\Psi_\gamma$ from that of the two halo neutrons $\Psi_{2n}$, one knows also this one, and the no return event (although most likely the $2n$ system long before has ended up as heat in the accelerator shielding). Namely, the falling of $2n$ into the no--return triton potential leading to $\Psi_t=f(\Psi_p,\Psi_{2n})$ ($\Psi_p$ describing the proton beam), and thus to its ($2n$) ultimate fate. Viceversa, observing $\Psi_t$ but not measuring neither the energy nor the momentum of the triton, provides complete information on $\Psi_\gamma$, and of the presence of a $\gamma$--ray of frequency $\nu_{2^+_1}$ and intensity $I_\gamma$. Whether it reaches the detector or ends up contributing to the (local) background radiation or detector shielding heat, is a question of detector budget. 

But the possibilities within the scenario of entanglement and correlation in nuclear structure and reactions are richer than anticipated above. In fact, by changing the bombarding energy of the proton one expects a resonant behaviour when the de Broglie wavelength matches a value related to the wavelength of $h\nu_{2^+}$ in each of the reactions considered (self energy processes Figs. \ref{fig5} (a) and (d), and Fig. \ref{fig6} (a)) but also tht of the dipole mode in the second one (vertex correction, induced pairing, Fig. \ref{fig6} (b)) which essentially provides all of the small but finite ($S_{2n}=380$ keV) energy, binding the two halo neutrons to the core $^9$Li. By making the proton beam oscillate between the differential cross sections resonant behaviour bombarding energies associated with $h\nu_{2^+}$ and $h\nu_{1^-}$, one would mimic a kind of self--amplifying Hawking radiation. Technical difficulties likely restricts this to remain an only  \textit{gedanken eksperiment}.

Nonetheless, in the nuclear case, there are further degrees of entanglements. At the level of nuclear structure, in keeping with the fact that the bosonic elementary modes of excitation are not elementary but composite two--\\quasiparticle--like collective excitations. At the level of nuclear reactions in which case two--nucleon transfer is completely dominated by successive transfer, due to the fact that the correlation energy of Cooper pairs is much smaller than the Fermi energy ($\approx10^{-2}$ in the case of $^{11}$Li), and that the correlation length between members of the pair is larger than nuclear dimensions.

Let us elaborate on these points, using as examples $\ket{^{11}\text{Li (gs)}}$, $\ket{^{11}\text{Li}(1^-;750\text{ keV})}$ and $\ket{^{9}\text{Li}(2^+;3.368\text{ MeV})}$ and the reaction $^{11}$Li$(p,t)^9$Li $(1^-$; 2.691 MeV). In all these states it is assumed that the $p_{3/2}(\pi)$ odd proton acts as a spectator and thus we do not write it for simplicity. We start with $\ket{^{11}\text{Li (gs)}}$, namely the two neutron halo pairing correlated system. Making use of the microscopic random phase approximation (RPA) and of the quasiparticle RPA (QRPA) description of $\ket{^{9}\text{Li}(2^+;3.368\text{ MeV})}$ and of $\ket{^{11}\text{Li}(1^-;750\text{ keV})}$ respectively, one can calculate the self--energy contributions and the induced pairing interaction (vertex corrections) shown in Figs. \ref{fig6} (a) and (b), using also the $v_{14}$ Argonne potential as the bare $NN$--$^1S_0$ bare pairing interaction. Propagating these processes to infinite order by solving a Dyson--like equation, one obtains an accurate description of the experimental findings (for details see \cite{Broglia:16} and \cite{Potel:10}).

\begin{figure*}
 	\begin{center}
 		%\fbox{\includegraphics[width=0.7\textwidth]{fig10.pdf}}
	 \includegraphics[width=0.8\textwidth]{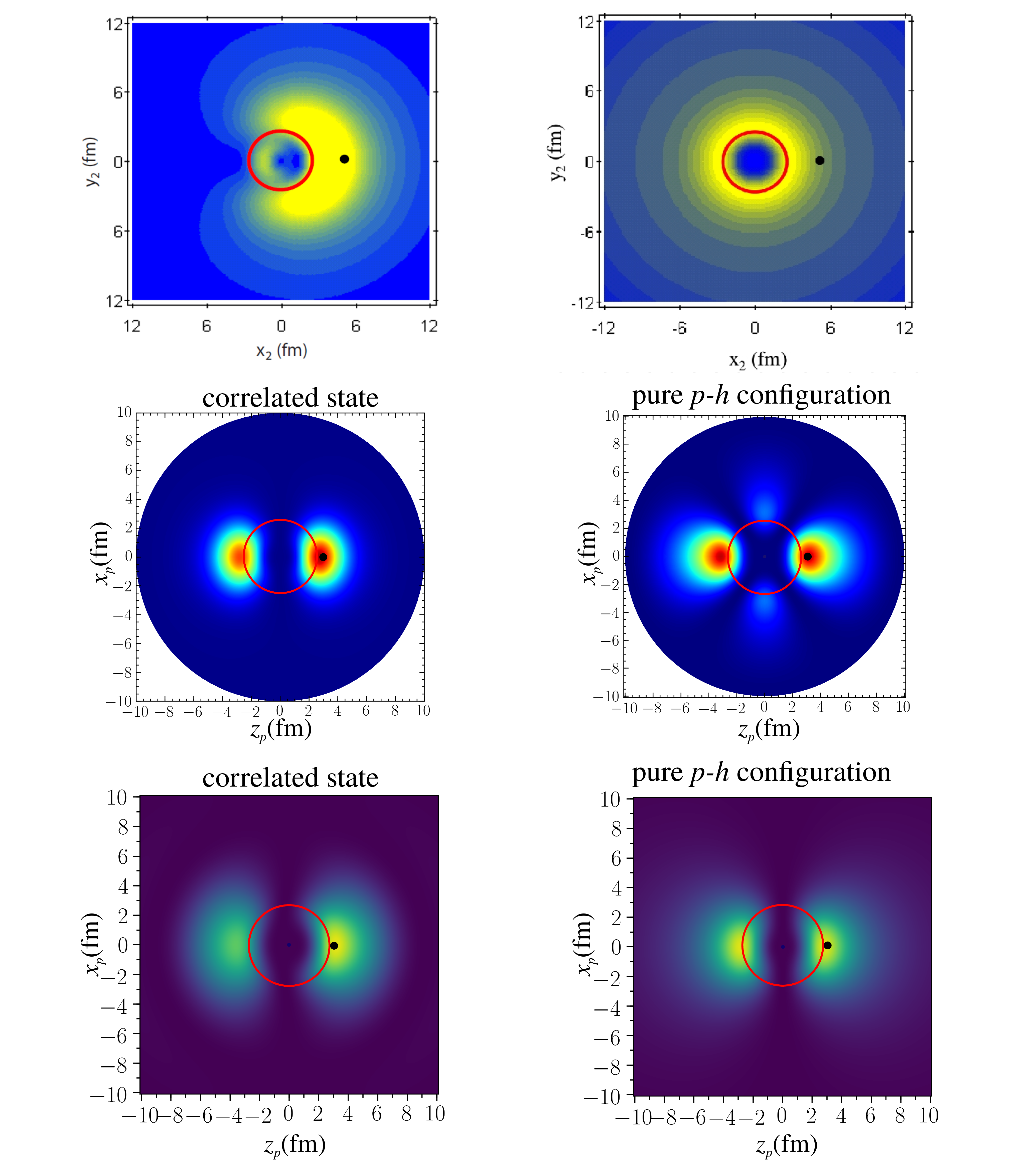}
 	\end{center}
 	\caption{ (\textbf{Upper left}) Spatial structure of the neutron halo Cooper pair $\ket{\tilde 0}_\nu$entering in the $^{11}$Li ground state $\ket{\tilde 0_\nu}\otimes\ket{p_{3/2}(\pi)}$, the proton assumed to act as a spectator. The modulus squared of the wavefunction $\Psi_0(\mathbf r_1,\mathbf r_2)=\braket{\mathbf r_1,\mathbf r_2|\tilde 0}_\nu$ is displayed as a function of the Cartesian coordinates $x_2,y_2$ of particle 2, for a fixed position of particle 1 ($x_1=5$ fm, $y_1=0$; solid dot). The red circle represents the radius of $^9$Li $R_0=2.5$ fm. (\textbf{Upper right}) same, but for the uncorrelated pure configuration ($p_{1/2})^2$. (\textbf{Middle left}) same as above, but for the ($p-h$) quadrupole vibration of $^{10}$Be. In this case the results are displayed in terms of the Cartesian coordinates ($x_p,z_p$) of the particle for a fixed position of the hole ($z_h=3$ fm, $x_h=0$ fm; solid dot). The red circle represents the radius of $^{10}$Be $R_0=2.6$ fm. (\textbf{Middle right}) same, but for the neutron uncorrelated particle--hole configuration $\left((s_{1/2})^{-1},d_{5/2}\right)_{2^+}$. (\textbf{Down left}) Same as above (middle) but for the case of the soft dipole mode of $^{11}$Li. The radius is equal to that of the upper figure (\textbf{Down right}) same as above but for the uncorrelated neutron particle--hole configuration $\left((p_{1/2})^{-1},s_{1/2}\right)_{1^-}$.
 	}\label{fig7bis}
 \end{figure*}

In Fig. \ref{fig7bis} (a) and (b) we display the resulting spatial correlations of the two neutrons in $\ket{^{11}\text{Li (gs)}}$ and compare it with the pure configuration $1p^2_{1/2}(0)$, an important component of the ground state of $^{11}$Li. Similar results are shown in connection with the $2^+$ of $^9$Li and the $1^-$ of $^{11}$Li (Figs. \ref{fig7bis} (c) and (d) and (e), (f)). The importance of the correlation is apparent.

Let us concentrate now on entanglement regarding the two--nucleon transfer process. As seen in Fig. \ref{fig6} (d), the transfer of one nucleon at a time, that is successive transfer, constitutes the main contribution to the transfer process. From the particle--particle correlation displayed in Fig. \ref{fig7bis} (a) and (b), and the fact that the two neutrons in the triton are close by ($\sim 2$ fm) one would have expected simultaneous transfer to be the main component. Now, the probability of neutron tunneling decreases exponentially with the square root of the mass. Because pairing correlations have a coherence length larger than nuclear dimensions, it is thus profitable that one nucleon tunnels at a time. Said it differently, to calculate the probability $P_2$  of a two--particle transfer process of a pair of correlated nucleons, one has to add the phased single--particle tunneling probability amplitudes, before taking the absolute square value, that is
\begin{align}
P_2=\left|\frac{e^{i\phi_1}\sqrt{P_1}+e^{-i\phi_2}\sqrt{P_1}}{2}\right|^2=\frac{P_1}{2}(1+\cos(\phi_1+\phi_2)),
\end{align}
and thus $P_2\approx P_1$ $(\phi_1+\phi_2\approx 0)$, a result which parallels that found by Anderson \cite{Anderson:64b} in connection with the Josephson effect. Typical examples of $P_2\approx P_1$ in the nuclear case are provided by \cite{Tanihata:08}\\ $d\sigma(^9\text{Li} (d,p)^{10}\text{Li} (1/2^-))/d\Omega|_{\theta_{max}}\approx0.8$ mb/sr, as compared to \cite{Cavallaro:17}  $d\sigma(^{11}\text{Li} (p,t)^{9}\text{Li} (1/2^-))/d\Omega|_{\theta_{max}}\approx1 $ mb/sr, \cite{Fortune:94} $^{10}$Be$(t,p)^{12}$Be(gs) $(\sigma=1.9\pm0.5$ mb, $4.4^\circ\leq\theta_{cm}\leq54.4^\circ)$ as compared to \cite{Schmitt:13} 
$^{10}$Be$(d,p)^{11}$Be($1/2^+$) $(\sigma=2.4\pm0.013$ mb, $5^\circ\leq\theta_{cm}\leq39^\circ)$ in the case of light nuclei around closed $(N=6)$ shell, and \cite{Bassani:65} $^{120}$Sn$(p,t)^{118}$Sn(gs) $(\sigma=3.024\pm0.907$ mb, $5^\circ\leq\theta_{cm}\leq40^\circ)$ as compared to \cite{Bechara:75} $^{120}$Sn$(d,p)^{121}$Sn($7/2^+$) $(\sigma=5.2\pm0.6$ mb, $2^\circ\leq\theta_{cm}\leq58^\circ)$.

Let us now discuss the correlation between particles and holes ($ph$) associated with the two quasiparticle ($pp,ph,hh$) states $2^+_1$ and $1^-$, and at the basis of the phenomena of core polarization responsible for the dressing of particles (self energy) and the renormalization of the $ph$ and ($pp,hh$) interactions (vertex and pairing renormalization).

As seen from Figs. \ref{fig7bis} (c), (d) and (e), (f), the ($ph$) become closer together when correlated by the quadrupole and the dipole residual interaction, respectively. Emitted and reabsorbed by single nucleons (Fig. \ref{fig3}, Fig. \ref{fig5} I (a), (d), Fig. \ref{fig6} (a)) they give rise to the quasiparticle degrees of freedom carrying effective masses (energies) and spectroscopic amplitudes (single--particle content), as experimentally observed (see e.g. \cite{Barranco:17,Barranco:19} and refs. therein). Exchanged between nucleons they renormalise the bare nucleon--nucleon interaction, in particular the $^1S_0$ pairing interaction (see e.g. \cite{Barranco:99,Terasaki:02a} and refs. therein), effects which can be treated in nuclear field theory also to infinite order of perturbation if needed, in particular in the case of superfluid nuclei, but also of halo nuclei like $^{11}$Li. Ground state correlations of $ph$ collective modes and associated renormalization effects provide non negligible contributions to the binding energies (see Figs. \ref{fig5} (II) (a)--(c), Table \ref{table3} and e.g. \cite{Baroni:04} and refs. therein). They are also essential in reproducing the experimental value of the electromagnetic transition probabilities. In fact, dressing the collective vibrations, e.g. the collective quadrupole mode of $^{120}$Sn, leads to conspicuous increase in the $B(E2)$--value associated with the decay into the ground state \cite{Barranco:04}, in overall agreement with the experimental findings.

 Clearly, this can hardly be connected with whether particles and holes are close in space, as the wavelength of $\gamma$--rays of 1--2 MeV are orders of magnitude larger than nuclear dimensions. In fact, it is related to the fact that the components of the wavefunctions of collective states are phase--correlated, as is the case  in $pp$ correlated states (pairing vibrations like $\ket{^{11}\text{Li (gs)}}$) and associated two--particle, mainly successive, transfer.
 
\begin{figure*}
    	\begin{center}
    	\centerline{\includegraphics*[width=11cm,angle=0]{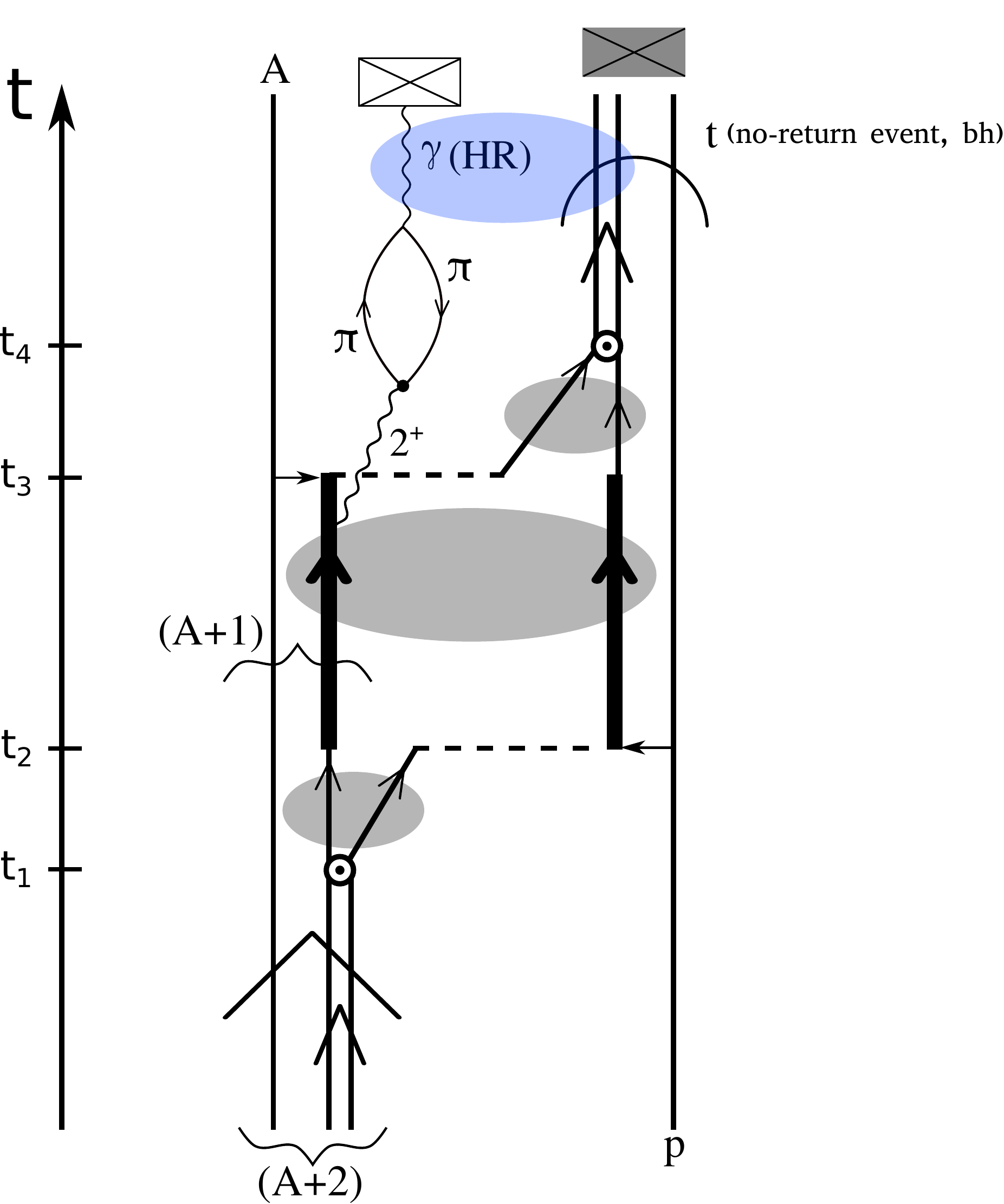}}
    	\end{center}
    	\caption{General nuclear field theory diagram describing structure and reaction aspects of the main process through which a Cooper pair (di--neutron) tunnels from target to projectile in the reaction $(A+2)+p\to A+t$. In order that the two--step process $(A+2)+p\to(A+1)+d\to A+t$ takes place, target and projectile have to be in contact at least in the time interval running between $t_2$ and $t_3$. During this time, the two systems create, with local regions of ever so low nucleonic presence, a common density over which the non--local pairing field can be established, and the Cooper pair can be correlated. Even with regions in which the pairing interaction may be zero. Small grey ellipses (with linear dimensions of the order of the nuclear radius $R_0$) indicate situations in which the two neutron correlation is distorted by  the  mean field of a single of the systems involved of the reaction, i.e. $A+2$ in the entrance channel, $t$ in the exit one. Mean field which can be viewed as acting as an external field. The large grey ellipse (with linear dimensions of the order of $\xi$) indicate the region in which the two partners of the Cooper pair correlate over distances of the order of the correlation length. It is this information that the outgoing particle of a Cooper pair transfer process brings to the detector. In other words, this is the closest to what can be defined as the observable Cooper pair in terms of its specific probe, i.e. two--nucleon transfer process, and the reason why the neutrons are described, in the interval $\Delta t=t_3-t_1$, in terms of bold face arrowed lines. In the present case the diagram is tailored to describe the process shown schematically in Fig. \ref{fig6} (b), i. e. $^{11}$Li$(p,t)^9$Li($1/2^-$). Namely a situation in which the irreversible event associated with the tunneling of the second fermion takes place before the collective vibration is either reabsorbed by it, or is exchanged with the first neutron, forcing the collective mode to become on-shell and, after coupling to the electromagnetic field,  escape the reaction area, entangled (large light blue ellipse) with the two fermions (neutrons) which have fallen into the no-return field of the triton (see Fig. \ref{fig1} (I) (b)).}\label{fig8}
    \end{figure*}

A summary of correlation and entanglement simultaneously operative at the level of structure and reaction discussed above, is shown in Fig. \ref{fig8}, for the case of the process $^{A+2}X+p\to^{A}X(J^\pi;E_x)+t$ (e.g $^{11}$Li+$p\to^9$Li($1/2-;2.69$ MeV)+$t$). The small (grey) ellipses focus on the particle--particle (neutron--neutron) correlations. That is, a structure property which is calculated for the systems ($A+2$) and $t$ ($\equiv^3$H) in isolation. The corresponding wavefunctions describe the effect of both ($pp$)--correlation (weak), as well as that of the external single--particle field (strong). The large ellipse focus on the $(pp)$ entanglement taking place in the transfer process, dominated by the mechanism of successive transfer. The outgoing triton and $\gamma$--ray (resulting from the $E2$--decay of the quadrupole mode of the core $A$ ($^9$Li) are entangled and bring the specific information regarding the correlation existing between the fermionic partners of the Cooper pair, closely connected with the transfer formfactor. The $(p,t)$ is an irreversible, no--return process providing the energy, momentum both linear and angular for the $\gamma$ ray to become on shell. The variety of processes are treated fully quantum mechanically and on equal footing, within the full single--particle space, described by both bound and continuum states.

\section{Conclusions}

The vacuum state of a quantal system contains,  through zero point fluctuations, virtual
information concerning the particles (elementary modes of excitation in 
the case of a many-body system)  building the system, and  their interactions (interweaving). 
To bring this information to the detector, one needs to intervene the virtual states, in a no--return fashion, with
external fields which share the properties one wants to observe.  In the nuclear case, one-particle 
transfer to learn about single-particle motion,  Cooper pair transfer 
to get information concerning  the mechanisms by which gauge invariance can be violated  (Cooper pair binding). 
Doing so in the case of light halo nuclei we have learned that, in a similar way in which the Lamb shift provided 
in the H-atom a definitive answer to Rabi's question of whether the polarisation 
of a QED vacuum could be measured,
%\cite{Pais}, 
parity inversion in nuclei provides a definitive 
answer of the central role collective vibrations play both  in the dressing processes of valence 
nucleons, as  well as in the induced pairing interaction acting among them, as testified by the Hawking-like radiation observed in
the $^1$H($^{11}$Be,$^{10}$Be($2^+$))$^2$H
and  $^1$H($^{11}$Li,$^9$Li($1/2^-$))$^3$H reaction
processes, respectively.
  To be able to recover  information contained in the vacuum associated with the 
field theoretical description of  the nuclear structure $(s)$, NFT had to be extended 
to be able to describe also reaction processes $(r)$ to the same  level of accuracy, and making use of the same language \cite{Broglia:16}. In particular, treating  on equal footing  
non-orthogonality  and non-locality of the elementary modes of excitation as well as
simultaneous, successive and non-orthogonality (non-local-)contributions to Cooper pair tunnelling. Within this context,
one can refer to the simultaneous renormalisation
of single-particle energies and transfer form factors as a further consequence of the 
above $(s+r)$ unification requirement.

Say it differently, we have critically assessed  experimental information shedding light on one- and two-neutron halo nuclei 
$ ^{11}$Be \cite{Winfield:01} and $^{11}$Li  \cite{Tanihata:08} respectively, discuss the texture of the associated nuclear vacuum
and point to possible nuclear parallels to Hawking radiation in the sense of the abstract. It is of notice that 
in spite of much effort, a theory of quantum gravity which unifies general relativity and quantum mechanics
does not yet exist. The prediction of Hawking radiation results from a combination of these two theories.
In the nuclear case, a unified quantal description of structure and reactions (which play the role of the irreversible, no-return event), taking into account nonlocality and retardation
both concerning correlations and transfer mechanisms and involving  also electromagnetic 
decay, is available within the framework of renormalised NFT \cite{Broglia:16}.

 It is our hope that the subjects discussed in the present paper, presenting a new view of nuclear dynamics in connection with black hole Hawking radiation can act as intellectual stimulus concerning a full quantum treatment of the phenomena involved. From our part we consider this the first step of a major challenge which we plan to follow up in future publications.

 \section{Acknowledgment}
R.A.B.  is grateful to  L. Mandelli for discussions as well as to C. Pethick for suggestions. 
F.B. and E.V. acknowledge funding from the European Union Horizon 2020 research and innovation program,
 under Grant Agreement No. 654002. F.B. acknowledges funding from the Spanish
Ministerio de Econom\`\i a  under Grant Agreement FIS2014-53448-C2-1-P.

\appendix 
\section{Feynman rules for  calculating the S-matrix in QED}

\begin{figure*}
 	\begin{center}
 		%\fbox{\includegraphics[width=0.7\textwidth]{fig10.pdf}}
 		\includegraphics[width=0.8\textwidth]{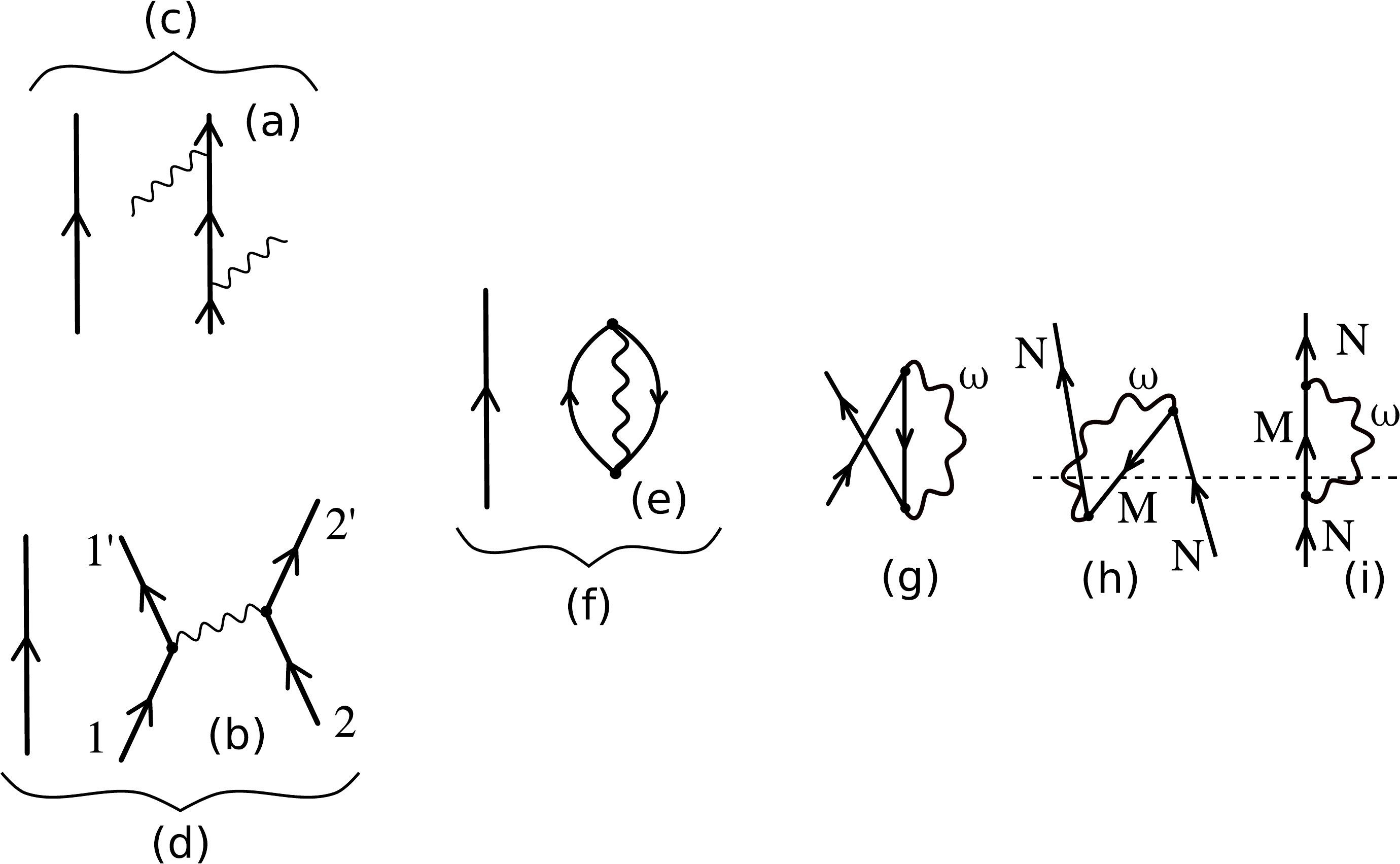}
 	\end{center}
 	\caption{ {\bf (a)} Lowest -order Feynman diagram for Compton scattering.
 		{\bf (b)} Scattering of two charged particles. Lines carrying arrows pointing upwards 
 		are electrons, downwards positrons, wavy lines being photons. At each vertex $V(t)$ (Eq. (\ref{eq11}))
 		is operative. Time is assumed to run upwards.
 		{\bf (c)} and  {\bf (d)}, same as { (a)} and {(b)} but with a disconnected 
 		electron line (spectator). 
 		{\bf (e)} Joining two wavy lines, and the two electron lines of (a), or positron-electron 
 		lines (2 with 1, and 1' with 2') in (b)  leads to the lowest-order, two vertex, vacuum fluctuation 
 		process.
 		{\bf (f)} Same as (e) but with the spectator electron line.
 		{\bf (g)} Process resulting from the exchange of the two electron lines.
 		{\bf (h)} Identical process to (g) redrawn to keep with standard  presentation. 
 		{\bf (i)} electron self-energy process resulting from the time ordering rearrangement of (h).
 	}\label{fig7}
 \end{figure*}

One considers electrons, positrons and photons. A possible gauge- and Lorentz-invariant Lagrangian 
for QED is 
\begin{equation}
{\cal L} = -\frac{1}{4} F_{\mu\nu}F^{\mu\nu}- \bar{\Psi}(\gamma^{\mu}[\partial_{\mu} + i e A_{\mu}]+m)\Psi
\end{equation}
where 
\begin{equation}
F_{\mu\nu}(x) = \partial_{\mu}A_{\nu}(x)- \partial_{\nu}A_{\mu}(x),
\end{equation}
$m$ being the electron mass, $\gamma^{\mu}$ Dirac matrices and one sums
over indices like $\mu$ and $\nu$ which appear twice, one upstairs and the other downstairs. The
electric current four-vector is 
\begin{equation}
J^{\mu} = \frac{\partial \cal L}{\partial A_{\mu}} = -i e \bar{\Psi} \gamma^{\mu}\Psi.
\end{equation}
The interaction is
%\begin{equation}
%V = - \int d^3 x \vec J \cdot \vec A + V_{coul},
%\end{equation}
%in the interaction picture is
\begin{equation}\label{eq11}
V(t) =   i e \int d^3x (\bar{\Psi}(\vec x,t) \gamma^{\mu}\Psi(\vec x,t)) a_{\mu}(\vec x,t) + V_{coul},
\end{equation}
where electrons are created and annihilated by fields ${\bar\Psi}$ and $\Psi$, while the photon is created and annihilated
by fields $a_{\mu}$, the instantaneous Coulomb field $V_{coul}(t)$ just serving to cancel the part of the photon 
propagator that is non-covariant and local in time, and 
\begin{equation}\label{eq12}
\mathbf{A}(\mathbf x)=\sum_{\lambda=\pm}\int \widetilde{dk}\left[\epsilon^*_\lambda(\mathbf k)a_\lambda(\mathbf k) e^{ikx} + \epsilon_\lambda(\mathbf k)a^\dagger_\lambda(\mathbf k)e^{-ikx}\right],
\end{equation}
where $k^0=\omega=|\mathbf k|$, $\widetilde{dk}=d\mathbf k/2(\pi)^3 2\omega$ and $\epsilon_+$ and $\epsilon_-$ are polarization vectors.
Let us now remind some of the Feynman rules for calculating the connected part of the $S-$matrix of quantum 
electrodynamics. 
In particular the first one: (i) draw all Feynman diagrams with up to some number of vertices. The diagrams consist
of electron lines carrying arrows, with the lines joined at the vertices, a each of which there is one incoming and one outgoing 
electron line and one photon line.
Examples of diagrams containing two vertices are provided by lowest-order Compton scattering 
and electron-positron scattering (Fig. \ref{fig7} (a) and (b), respectively).

Another two-vertex diagram is obtained by joining in graph {\bf (a)} the photon lines and the electron lines,
in this last case going backwards in time, and the  1' - 2' and $1-2$ electron line  in {\bf (b)}. The resulting
diagram {\bf (e)} describes the lowest order  vacuum fluctuation.

In drawing these diagrams one should exclude disconnected diagrams, that is, diagrams in which any $V(t)$ operator or any
initial or final particle is not connected to every other one by a sequence of particle creations and annihilations.
Examples are provided by diagrams {\bf (c),(d)} and {\bf (f)} of Fig. \ref{fig7}.

If one views the disconnected electron line in {\bf (f)} as describing  the electron of an hydrogen atom,
virtual fluctuations can affect the energy levels, in particular that of the $2s_{1/2},2p_{1/2}$ levels predicted by Dirac 
equation to have the same energy.
This is because the electron in the zero point fluctuation (ZPF) of the vacuum may partially occupy the same state
occupied by the electron of the hydrogen atom.

The exchange of the virtual and the disconnected electron lines correct for Pauli principle  violation, leading to 
the connected two-vertex diagram {\bf (g)} (identical to {\bf (h)}) and, by time ordering, to diagram {\bf(i)},
The energy denominators, associated with these diagrams, difference of the initial and of the intermediate 
state energy are $Den_{(h)}= -  (E_N +E_M +\omega)$  and $Den_{(i)}= [E_N - (E_M + \omega$)]  respectively.
In these expressions  $E_N$ is the energy of the electron in the initial state, the energy of the intermediate state
being either the sum of the energy $E_M$ of an electron and a photon of energy $\omega$ (Fig. \ref{fig7} (i)),
or else a positron of energy $E_M$, a photon of energy $\omega$, plus $2E_N$, sum of the energy of both initial and 
final electrons (Fig. \ref{fig7} (h)).

The resulting value  of the difference between the summed contributions $(h)+(i)$ associated with the
$2s_{1/2}$ and $2p_{1/2}$ states of the hydrogen atom from relativistic calculations leads to
$[\delta E]_{2s} - [\delta E]_{2p_{1/2}} = 1052.19 $ MHz  \cite{Weinberg:96} (see also \cite{Welton:48,Kroll:49,French:49}, and \cite{Bethe:47}) 
as compared to the experimental
value 1057.845(9) MHz \cite{Lundeen:81}, the value reported by the first experiment  and  (non-relativistic)
theoretical calculations being 1000  MHz \cite{Lamb:47} and 1040 MHz respectively \cite{Bethe:47}.
As exemplified  by the Feynman  diagrams shown in Fig. \ref{fig7}, by construction  and as a direct 
consequence of the interaction (A.4) which is bilinear in electron fields and linear in photon fields, 
associated with each vertex  there are two electron (positron) lines an one photon line. 
In particular in the case of the process shown in Fig. \ref{fig7} (e), namely  a two vertex Feynman diagram
describing the zero point fluctuations of the QED vacuum. 

As recounted by Pais \cite{Pais:86,Pais:00}, Lamb provided a quantitative answer, both experimentally
and theoretically \cite{Kroll:49,Lamb:47} to the question  of Rabi of whether the polarisation of the vacuum could be measured. 
According to quantum mechanics, intervening a virtual process as a result of a conservation law 
or a physical principle (external field), in the present case the exclusion principle, one can probe 
the structure of the associated off-the-energy-shell process. Within this context and of the Lamb shift phenomenon,
the need of a three lines, two electrons (positrons) and one photon, virtual process in connection 
with QED vacuum fluctuations is apparent. 

It could be argued that an electron-positron vacuum virtual excitation can be a valid QED Feynman 
diagram interpreting the vertices as  the result of the action of  a Coulomb field.  As stated in 
connection  with Eq. (A.5), the violation of Lorentz invariance by the instantaneous Coulomb interaction, 
is cancelled by another apparent violation of Lorentz invariance connected  with the fact that
the photon fields $a_{\mu}$ are not four vectors, and therefore have a  non-covariant propagator. 
The important point is that the photon propagator is taken effectively as a covariant quantity 
\begin{equation}
\Delta^{eff}_{\mu\nu}(x-y) = (2 \pi)^{-4} \int d^4 q \frac{\eta_{\mu\nu}}{q^2 -i \epsilon} e^{iq(x-y)}
\end{equation}
with the Coulomb interaction dropped.
From a practical point of view, the main issue s that in the momentum space Feynman rules, the contribution
of an internal photon line is given by 
\begin{equation}
\frac{-i}{(2 \pi)^4} \frac{\eta_{\mu\nu}}{q^2 - i \epsilon},
\end{equation}
and the Coulomb interaction dropped, as reflected by (A.5) (for details see \cite{Weinberg:96}).

%\bibliographystyle{unsrt}
%\bibliography{/home/gregory/book/nuclear_bib}

 \end{document}